\documentclass[a4paper,conference]{IEEEtran}

\IEEEsettopmargin{t}{30mm}
\IEEEquantizetextheight{c}
\IEEEsettextwidth{14mm}{14mm}
\IEEEsetsidemargin{c}{0mm}

\usepackage[cmex10]{amsmath} 
\usepackage{sty/style}

\def\BibTeX{{\rm B\kern-.05em{\sc i\kern-.025em b}\kern-.08em
    T\kern-.1667em\lower.7ex\hbox{E}\kern-.125emX}}
\begin{document}


\title{On the Complexity of \\ Electromagnetic Far-Field Modeling}


\author{%
  \IEEEauthorblockN{Torben K\"olle, Alexander Stutz-Tirri, and Christoph Studer}\\
  \IEEEauthorblockA{\em Department of Information Technology and Electrical Engineering, ETH Zurich, Switzerland \\ email: tkoelle@ethz.ch, alstutz@ethz.ch, studer@ethz.ch}
  \thanks{The work of TK, AST, and CS was funded in part by the Swiss State Secretariat for Education, Research, and Innovation (SERI) under the SwissChips initiative, by a CHIST-ERA grant for the project CHASER (CHIST-ERA-22-WAI-01) through the SNSF grant 20CH21 218704, and by the European Commission within the context of the project 6G-REFERENCE (6G Hardware Enablers for Cell Free Coherent Communications and Sensing), funded under EU Horizon Europe Grant Agreement~101139155. The work of AST was funded in part by armasuisse.}
}

\maketitle


\begin{abstract}   
Modern wireless systems are envisioned to employ antenna architectures that not only transmit and receive electromagnetic (EM) waves, but also intentionally reflect and possibly transform incident EM waves. In this paper, we propose a mathematically rigorous framework grounded in Maxwell’s equations for analyzing the complexity of EM far-field modeling of general antenna architectures. We show that---under physically meaningful assumptions---such antenna architectures exhibit limited complexity, i.e., can be modeled by finite-rank operators using finitely many parameters. Furthermore, we construct a sequence of finite-rank operators whose approximation error decays super-exponentially once the operator rank exceeds an effective bandwidth associated with the antenna architecture and the analysis frequency. These results constitute a fundamental prerequisite for the efficient and accurate modeling of general antenna architectures on digital computing platforms.
\end{abstract}

\section{Introduction}\label{sec:introduction}
In conventional wireless communication systems, antenna elements are typically only used for transmitting and receiving electromagnetic (EM) waves---rather than reflecting EM waves incident from outside the system. 
In such systems, the far-field interaction\footnote{With far-field interaction, we refer to the interaction with EM waves (i)~that are incoming from sufficiently far away to be represented as convergent spherical waves and (ii) that are outgoing into the far-field region.} 
admits an exact finite-rank description~\cite[Sec.~II-B4]{stutz_schwan_studer_efficient_and_physically_consistent_modeling_of_reconfigurable_electromagnetic_structures}.\footnote{Specifically, the far-field interaction can be characterized by the finite-rank operators~$\oper{S}_{\up{R}_\up{RR}}$, $\oper{S}_{\up{R}_\up{FR}}$, and $\oper{S}_{\up{R}_\up{RF}}$ introduced in~\cite[Eq.~17]{stutz_schwan_studer_efficient_and_physically_consistent_modeling_of_reconfigurable_electromagnetic_structures}.} 
Consequently, conventional wireless systems can, in principle, be modeled using finitely many parameters. 
In recent years, more general wireless architectures---such as array lenses~\cite{venkatesh_lu_saeidi_sengupta_a_high_speed_programmable_and_scalable_terahertz_holographic_metasurface_based_on_tilded_cmos_chips} and reconfigurable intelligent surfaces (RISs)~\cite{di_renzo_zappone_debbah_alouini_smart_radio_environments_empowered_by_reconfigurable_intelligent_surfaces,wu_zhang_zheng_you_zhang_intelligent_reflecting_surface_aided_wireless_communications}---have gained popularity. 
For such architectures, the reflection (or transmission) of EM waves incident from outside the system is crucial. 
Therefore, the question arises whether the far-field interaction in such modern wireless system architectures can still be modeled exactly---or at least approximated arbitrarily well---using finitely many parameters.
If this holds true, then we speak of \emph{limited complexity}. 
Limited complexity is a fundamental prerequisite for implementing models of wireless systems on digital computing platforms.
\subsection{Contributions}

In this paper, we study the complexity of wireless system modeling through the lens of linear system theory. 
We analyze whether a system’s external interaction is of limited complexity. 
In particular, we study whether a system can be approximated arbitrarily well by a sequence of finite-rank operators, i.e., whether it can, in principle, be modeled using finitely many parameters. 
To this end, we propose a mathematically rigorous framework grounded in Maxwell’s equations to analyze the modeling complexity of far-field interactions in wireless systems. 
With this framework, we prove that the far-field interaction of a large class of antenna systems is indeed of limited complexity---this property is established in~\fref{thm:finite_rank_approximation_far_field}. 
Furthermore, we explicitly show how to construct a sequence of finite-rank operators whose approximation error exhibits super-exponential decay once the rank exceeds an effective bandwidth associated with the antenna architecture and the analysis frequency---this property is established in~\fref{thm:vsh_approximation_far_field}.

\subsection{Prior Art}
In references~\cite{bucci_franceschetti_on_the_degrees_of_freedom_of_scattered_fields,bucci_franceschetti_on_the_spatial_bandwidth_of_scattered_fields}, Bucci and Franceschetti showed that scattered EM fields exhibit bandlimited-like behavior. 
Specifically, they proved that, sufficiently far away from a scatterer, (i) the scattered field can be uniformly approximated by a finite sum of basis functions, (ii) the corresponding approximation error exhibits a step-like dependence on the number of basis functions used, and (iii) the threshold beyond which the error rapidly decreases---i.e., the effective spatial bandwidth---depends on the scatterer’s electrical size.
However, these results quantify the complexity of the scattered fields rather than that of the scattering operator itself, and they are limited to electrically large scatterers. 
In contrast, we present results that are directly related to the system’s complexity, rather than the complexity of the system’s output, and our theorems are not restricted to large systems. 
In references~\cite{coifman_rokhlin_wandzura_the_fast_multipole_method_for_the_wave_equation_a_pedestrian_prescription,kalfa_erturk_ergul_error_analysis_of_mlfma_with_closed_form_expressions,song_chew_error_analysis_for_the_truncation_of_multipole_expansion_of_vector_greens_functions}, the authors analyze the approximation properties of Green’s functions. However, references~\cite{coifman_rokhlin_wandzura_the_fast_multipole_method_for_the_wave_equation_a_pedestrian_prescription,kalfa_erturk_ergul_error_analysis_of_mlfma_with_closed_form_expressions} are restricted to the scalar Green’s function. Moreover, reference~\cite{coifman_rokhlin_wandzura_the_fast_multipole_method_for_the_wave_equation_a_pedestrian_prescription} does not discuss the step-like behavior of the approximation error, whereas~\cite{kalfa_erturk_ergul_error_analysis_of_mlfma_with_closed_form_expressions,song_chew_error_analysis_for_the_truncation_of_multipole_expansion_of_vector_greens_functions} analyze the location of such a step-like transition; however, their analysis is limited either to small systems or to large systems. In addition, all three works focus on the approximation error of the scalar or dyadic Green’s function, rather than on the resulting error in the EM fields. 
In contrast, without restricting ourselves to small or large systems, we (i) directly analyze the complexity of wireless systems, and (ii) identify and characterize where the step-like approximation error transition occurs. 
In reference~\cite{koivisto_analytical_solution_for_characteristic_modal_power_distribution_and_truncation_limit_for_spherical_wave_expansion_of_antenna_radiation_pattern}, Koivisto analyzes an expansion of electromagnetic fields radiated by an antenna. In particular, similarly to our framework, the fields are represented using spherical-harmonic basis functions.
However, that paper does not provide a general expression for the effective bandwidth applicable to general systems. 
In contrast, in~\fref{thm:vsh_approximation_far_field}, we provide an equation for the effective bandwidth of general systems. 

\subsection{Notation}

We use lowercase boldface for general vectors (e.g.,~$\vect{a}$) and uppercase boldface for general matrices (e.g., $\mat{A}$). 
We use pink sans-serif (e.g.,~$\phs{a}$) and pink sans-serif boldface (e.g.,~$\phv{a}$) for phasors (cf.~\cite[Def.~1]{stutz_schwan_studer_efficient_and_physically_consistent_modeling_of_reconfigurable_electromagnetic_structures}) and vectors containing phasors, respectively. 
The superscripts $^\T$ and $^\He$ indicate transpose (e.g.,~$\mat{A}^\T$) and conjugate transpose (e.g.,~$\mat{A}^\He$), respectively.
We denote the Euclidean norm by~$\|\cdot\|_2$. 
We use blackboard bold for operators (e.g.,~$\mathbb{S}$). 
Given an operator~$\oper{S}$, we denote its range by~$\range\{\oper{S}\}$ and its operator norm by~$\|\oper{S}\|_{\up{op}}$. 
Given two operators~$\oper{A}$ and $\oper{B}$, we use~$\oper{A}\circ\oper{B}$ to denote their composition. 
We denote the imaginary number by~$j\triangleq\sqrt{-1}$. 
For a complex number $z\in\mathbb{C}$, the conjugation is~$\overline{z}$. 
We use a calligraphic font for sets (e.g.,~$\mathcal{V}$), except for the sets of natural, integer, real, and complex numbers, denoted by $\mathbb{N}$, $\mathbb{Z}$, $\mathbb{R}$, and $\mathbb{C}$, respectively. 
Given $N\in\mathbb{N}$, we define the set~$[N]\triangleq\{1,\ldots,N\}$. 
Given the elements~$\{\vect{e}_k\}_k$ of a vector space, we use~$\spn\!\left\{\{\vect{e}_k\}_k\right\}$ to denote their span. 
Given a set~$\mathcal{V}$, we use~$L^2({\mathcal{V}},\mathbb{C}^3)$ to denote the Bochner space induced by the Lebesgue measure space on~${\mathcal{V}}$ and the canonical Hilbert space on~$\mathbb{C}^3$. 
Given a Hilbert space~$\mathcal{H}$ and~$\vect{a},\vect{b}\in\mathcal{H}$, we use~$\langle\vect{a},\vect{b}\rangle_\mathcal{H}$ to denote the inner product with linearity in the first argument, and we define the shorthand notation~$[\vect{a}]_\vect{b}\triangleq \langle\vect{a},\vect{b}\rangle_\mathcal{H}$. 
We use the physicist's convention for spherical coordinate systems~\cite{ISO_quantities_and_units_2_mathematics}, with $r$ as the radial distance, $\theta$ as the polar angle, and $\varphi$ as the azimuthal angle.
To simplify notation, we use~$\vect{r}$ and~$\hat{\vect{r}}$ to denote the spherical coordinates $(r,\theta,\varphi)$ and the angular coordinates $(\theta,\varphi)$, respectively. 
For each position $\vect{r}$, we denote the local orthogonal unit vectors in the directions of increasing $r$, $\theta$, and~$\varphi$ as~$\hat{\vect{r}}$,~$\hat{\vect{\theta}}$, and~$\hat{\vect{\varphi}}$, respectively. 
Finally, we denote the set of all angular coordinates by $\Omega\triangleq[0,\pi]\times[0,2\pi)$. 
Throughout this paper,~$f$ refers to the frequency to be analyzed,~$\mu_\up{0}$ to the free-space permeability, $\varepsilon_\up{0}$ to the free-space permittivity,~\mbox{$k\triangleq 2 \pi f\sqrt{\mu_\up{0}\varepsilon_\up{0}}$} to the free-space wavenumber, and~$Z_\up{0}$ to the free-space impedance.


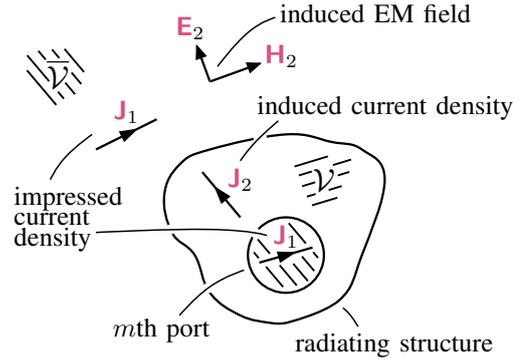
\begin{figure}[tp]
    \tikzstyle{every label}=[...]
    \centering
    {
    \begin{tikzpicture}
        \draw[physical] plot [smooth, samples=50, tension=0.7] coordinates {(0,-.9) (.7,-.6) (1,0) (1.3,1) (1,1.4) (.6,1.5) (0,1.6) (-1,1.4) (-1.4,1.2) (-1.6,.8) (-1.5,.3) (-1.2,0) (-1,-.5) (-.6,-.8) (-.2,-.9) (0,-.9)};
        \draw[physical] (0,0) circle (5mm);
        \node at (0,2.5mm) {$\phv{J}_1$};
        \draw[current] (-3.5mm,-1mm) -- (3.5mm,1mm);
        \begin{scope}[shift={(-5mm,-5mm)}]
             \node at (-16mm,24mm) {$\phv{J}_1$};
        \draw[current] (-20mm,19mm) -- (-12mm,23mm);
        \end{scope}
        \node at (-6mm,10mm) {$\phv{J}_2$};
        \draw[current] (-6mm,5mm) -- (-11mm,11mm);
        \begin{scope}[shift={(5mm,10mm)}, rotate=17]
            \draw[hatched] (-2.5mm,-2.8mm) -- (1mm,-2.8mm);
            \draw[hatched] (-2.8mm,-1.5mm) -- (-1.6mm,-1.5mm);
            \draw[hatched] (.2mm,-1.5mm) -- (2mm,-1.5mm);
            \draw[hatched] (-3mm,0mm) -- (-1.2mm,0mm);
            \draw[hatched] (1.9mm,0mm) -- (3.5mm,0mm);
            \draw[hatched] (-3mm,1.5mm) -- (-1.2mm,1.5mm);
            \draw[hatched] (1.7mm,1.5mm) -- (4.2mm,1.5mm);
            \draw[hatched] (-3.2mm,2.8mm) -- (1.4mm,2.8mm);
        \end{scope}
        \node at (5mm,10mm) {\large$\mathcal{V}$};
        \begin{scope}[shift={(0mm,0mm)}, rotate=130]
            \draw[hatched] (-3mm,-3mm) -- (-2mm,-3mm);
            \draw[hatched] (-.5mm,-3mm) -- (3mm,-3mm);
            \draw[hatched] (-4mm,-1.5mm) -- (-1.5mm,-1.5mm);
            \draw[hatched] (-4.5mm,0mm) -- (-1.5mm,0mm);
            \draw[hatched] (-4mm,1.5mm) -- (-.5mm,1.5mm);
            \draw[hatched] (1.5mm,1.5mm) -- (4mm,1.5mm);
            \draw[hatched] (-3mm,3mm) -- (.5mm,3mm);
            \draw[hatched] (2mm,3mm) -- (3mm,3mm);
        \end{scope}
        \node at (-30mm,24mm) {\large $\overline{\mathcal{V}}$};
        \begin{scope}[shift={(-30mm,24mm)}, rotate=130]
            \draw[hatched] (-2mm,-2.8mm) -- (2mm,-2.8mm);
            \draw[hatched] (-3mm,-1.5mm) -- (-1mm,-1.5mm);
            \draw[hatched] (1.5mm,-1.5mm) -- (3mm,-1.5mm);
            \draw[hatched] (2.5mm,0mm) -- (4mm,0mm);
            \draw[hatched] (-3.5mm,0mm) -- (-1.5mm,0mm);
            \draw[hatched] (-3mm,1.5mm) -- (-1.8mm,1.5mm);
            \draw[hatched] (0mm,1.5mm) -- (4.5mm,1.5mm);
            \draw[hatched] (-3mm,2.8mm) -- (3mm,2.8mm);
        \end{scope}
        \begin{scope}[shift={(-10mm,23mm)}, rotate=20]
            \draw[field_arrow] (0,0) -- (0.73,0);
            \draw[field] (0,0) -- (0.7,0);
            \draw[field_arrow] (0,0) -- (0,.53);
            \draw[field] (0,0) -- (0,.5);
            \node at (0,7.5mm) {$\phv{E}_2$};
            \node at (10mm,0mm) {$\phv{H}_2$};
        \end{scope}
        \begin{scope}[shift={(-37mm,8mm)}]
            \node[anchor=west] at (0,-.5mm) {impressed};
            \node[anchor=west] at (0,-3mm) {current};
            \node[anchor=west] at (0,-6mm) {density};
        \end{scope}
        \node[anchor=west] at (0mm,-12mm) {radiating structure};
        \node[anchor=west] at (-24mm,-10mm) {$m$th port};
        \node[anchor=west] at (-5mm,19.5mm) {induced current density};
        \node[anchor=west] at (-2mm,32mm) {induced EM field};
        \draw[label_line] plot [smooth, samples=10, tension=0.9] coordinates {(.95,-.95) (.85,-.8) (0.75,-.7)};
        \draw[fill=white,white] (-1.03,-.44) circle (.6mm);
        \draw[label_line] plot [smooth, samples=10, tension=0.9] coordinates {(-1.4,-.8) (-1,-.42) (-.55,-.2)};
        \draw[fill=white,white] (-.34,1.55) circle (.8mm);
        \draw[label_line] plot [smooth, samples=10, tension=0.9] coordinates {(-.1,1.78) (-.4,1.5) (-.6,1.25)};
        \draw[label_line] plot [smooth, samples=10, tension=0.9] coordinates {(-.14,3.15) (-.5,2.95) (-.9,2.5)};
        \draw[label_line] plot [smooth, samples=10, tension=0.9] coordinates {(-3.1,1) (-2.9,1.3) (-2.55,1.6)};
        \draw[fill=white,white] (-4mm,2.9mm) circle (.6mm);
        \draw[fill=white,white] (-1.52,3.4mm) circle (.6mm);
        \draw[label_line] plot [smooth, samples=10, tension=0.9] coordinates {(-2.5,.3) (-1,.34) (-2.5mm,2.6mm)};
    \end{tikzpicture}
    \caption{Analyzed problem setup: We consider a radiating structure that occupies the volume~$\mathcal{V}$, has~\mbox{$M$} ports, and is embedded in free space. Outside the structure’s volume, in the region~$\overline{\mathcal{V}}$, the current density~$\phv{J}_1$ is impressed. This excitation gives rise to an induced current density~$\phv{J}_2$ within the radiating structure, which in turn induces the electromagnetic field $(\phv{E}_2,\phv{H}_2)$.}
    \label{fig:setup}
    }
\end{figure}

\section{Problem Setup and Main Results}\label{sec:problem_setup_and_main_results} 
We consider general single- or multi-antenna systems of finite size. 
Specifically, we analyze \emph{radiating structures} for which~\fref{asm:finite_size} holds. 
\begin{defi}[Radiating Structure]\label{defi:radiating_structure}
    A \emph{radiating structure} is a passive physical object that interacts with the surrounding EM~field. 
\end{defi} 
\begin{asm}[Finite Size]\label{asm:finite_size}
    We assume that the analyzed radiating structure is of finite size. Concretely, there exists a finite radius~\mbox{$a\in\mathbb{R}_{\geq0}$} such that the radiating structure is completely contained within the ball 
    \begin{align} \label{eq:ball}
        \mathcal{B}_a
        &
        \triangleq
        \set{\vect{r}\in\mathbb{R}^3}
        {\|\vect{r}\|_2\leq a}\!.    
    \end{align}
\end{asm}
Given a radiating structure, we denote its physical volume by~$\mathcal{V}\subset\mathbb{R}^3$ and its surrounding region by $\overline{\mathcal{V}}\triangleq \mathbb{R}^3\setminus\mathcal{V}$. 
Furthermore, we fix an arbitrary analysis frequency $f\in\mathbb{R}_{>0}$ and focus on the associated single-frequency components. 
At this frequency, the excitation of the radiating structure can be characterized by an externally\footnote{Since radiating structures are passive by~\fref{defi:radiating_structure}, it is sufficient to consider only \emph{external} excitations.} impressed current density~\mbox{$\phv{J}_1\in L^2(\overline{\mathcal{V}},\mathbb{C}^3)$}. 
As depicted in~\fref{fig:setup}, we use~\mbox{$\phv{J}_2\in L^2(\mathcal{V},\mathbb{C}^3)$} to denote the current density induced in the radiating structure from~$\phv{J}_1$.
Moreover, we use~\mbox{$\phv{E}_2,\phv{H}_2\in L^2(\mathbb{R}^3,\mathbb{C}^3)$} to denote the EM field induced by~$\phv{J}_2$. 
\begin{figure}[tp]
    \tikzstyle{every label}=[...]
    \centering
    {
    \begin{tikzpicture}
        \draw[physical] plot [smooth, samples=50, tension=0.7] coordinates {(0,-.9) (.7,-.6) (1,0) (1.3,1) (1,1.4) (.6,1.5) (0,1.6) (-1,1.4) (-1.4,1.2) (-1.6,.8) (-1.5,.3) (-1.2,0) (-1,-.5) (-.6,-.8) (-.2,-.9) (0,-.9)};
        \draw[physical] (0,0) circle (5mm);
        \begin{scope}[shift={(5mm,10mm)}, rotate=17]
            \draw[hatched] (-2.5mm,-2.8mm) -- (1mm,-2.8mm);
            \draw[hatched] (-2.8mm,-1.5mm) -- (-1.6mm,-1.5mm);
            \draw[hatched] (.2mm,-1.5mm) -- (2mm,-1.5mm);
            \draw[hatched] (-3mm,0mm) -- (-1.2mm,0mm);
            \draw[hatched] (1.9mm,0mm) -- (3.5mm,0mm);
            \draw[hatched] (-3mm,1.5mm) -- (-1.2mm,1.5mm);
            \draw[hatched] (1.7mm,1.5mm) -- (4.2mm,1.5mm);
            \draw[hatched] (-3.2mm,2.8mm) -- (1.4mm,2.8mm);
        \end{scope}
        \node at (5mm,10mm) {\large$\mathcal{V}$};
        \begin{scope}[shift={(-11mm,-10mm)}]
            \node at (-30mm,24mm) {far-field region};
            \begin{scope}[shift={(-30mm,24mm)}, rotate=130]
                \draw[hatched] (-8.25mm,-9mm) -- (-3.75mm,-9mm);
                \draw[hatched] (-10mm,-7.5mm) -- (-8.5mm,-7.5mm);
                \draw[hatched] (-9mm,-6mm) -- (-7.25mm,-6mm);
                \draw[hatched] (-8mm,-4.5mm) -- (-6mm,-4.5mm);
                \draw[hatched] (-9mm,-3mm) -- (-5mm,-3mm);
                \draw[hatched] (-10mm,-1.5mm) -- (-3.5mm,-1.5mm);
                \draw[hatched] (-8.75mm,0mm) -- (-2.25mm,0mm);
                \draw[hatched] (-9mm,1.5mm) -- (-1mm,1.5mm);
                \draw[hatched] (-11mm,3mm) -- (0.25mm,3mm);
                \draw[hatched] (-13mm,4.5mm) -- (1.5mm,4.5mm);
                \draw[hatched] (-12mm,6mm) -- (2.75mm,6mm);
                \draw[hatched] (-10.5mm,7.5mm) -- (4mm,7.5mm);
                \draw[hatched] (-9.25mm,9mm) -- (5.25mm,9mm);
                \draw[hatched] (-7.75mm,10.5mm) -- (6mm,10.5mm);
                \draw[hatched] (-6mm,12mm) -- (4mm,12mm);
                \draw[hatched] (-4mm,13.5mm) -- (2mm,13.5mm);
                \draw[hatched] (-2mm,-7.5mm) -- (-4mm,-7.5mm);
                \draw[hatched] (-0.5mm,-6mm) -- (-2.5mm,-6mm);
                \draw[hatched] (1mm,-4.5mm) -- (-1.25mm,-4.5mm);
                \draw[hatched] (2.5mm,-3mm) -- (0mm,-3mm);
                \draw[hatched] (4mm,-1.5mm) -- (1.5mm,-1.5mm);
                \draw[hatched] (5mm,0mm) -- (2.75mm,0mm);
                \draw[hatched] (6.25mm,1.5mm) -- (4mm,1.5mm);
                \draw[hatched] (7.5mm,3mm) -- (5.25mm,3mm);
                \draw[hatched] (8.5mm,4.5mm) -- (6.25mm,4.5mm);
                \draw[hatched] (9.5mm,6mm) -- (7.5mm,6mm);
                \draw[hatched] (10mm,7.5mm) -- (8mm,7.5mm);
                \draw[hatched] (8.75mm,9mm) -- (7mm,9mm);
            \end{scope}
        \end{scope}
        \node[anchor=west] at (0mm,-12mm) {radiating structure};
        \node[anchor=west] at (-24mm,-10mm) {$m$th port};
        \draw[label_line] plot [smooth, samples=10, tension=0.9] coordinates {(.95,-.95) (.85,-.8) (0.75,-.7)};
        \draw[fill=white,white] (-1.03,-.44) circle (.6mm);
        \draw[label_line] plot [smooth, samples=10, tension=0.9] coordinates {(-1.4,-.8) (-1,-.42) (-.55,-.2)};
        \draw[physical,fill=white,white] (-3.8mm,3.2mm) circle (3.2mm);
        \begin{scope}[scale=.6,shift={(-12mm,8mm)},rotate=-45]
             \draw [field, shift={(0,0)}] plot[smooth, tension=.8] coordinates {(0,0) (.15,0) (.3,.025) (.45,-.1) (.6,.175) (.75,-.175) (.9,.1) (1.05,-.025) (1.2,0) (1.35,.0) };
            \draw [field, shift={(0,0)}, color=black, arrows = {-Stealth[inset=0, length=4pt, angle'=35]}] (0,0) -- (-0.1,0);
        \end{scope}
        \begin{scope}[scale=.6,shift={(1mm,3mm)},rotate=135]
             \draw [field, shift={(0,0)}] plot[smooth, tension=.8] coordinates {(0,0) (.15,0) (.3,.025) (.45,-.1) (.6,.175) (.75,-.175) (.9,.1) (1.05,-.025) (1.2,0) (1.35,.0) };
            \draw [field, shift={(0,0)}, color=black, arrows = {-Stealth[inset=0, length=4pt, angle'=35]}] (0,0) -- (-0.1,0);
        \end{scope}
        \node at (0,7.5mm) {$\phs{b}_m$};
        \node at (-7.5mm,1mm) {$\phs{a}_m$};
        \begin{scope}[shift={(-25mm,3mm)},rotate=43]
                \begin{scope}[scale=.6,shift={(-12mm,8mm)},rotate=-45]
                 \draw [field, shift={(0,0)}] plot[smooth, tension=.8] coordinates {(0,0) (.15,0) (.3,.025) (.45,-.1) (.6,.175) (.75,-.175) (.9,.1) (1.05,-.025) (1.2,0) (1.35,.0) };
                \draw [field, shift={(0,0)}, color=black, arrows = {-Stealth[inset=0, length=4pt, angle'=35]}] (0,0) -- (-0.1,0);
            \end{scope}
            \begin{scope}[scale=.6,shift={(1mm,3mm)},rotate=135]
                 \draw [field, shift={(0,0)}] plot[smooth, tension=.8] coordinates {(0,0) (.15,0) (.3,.025) (.45,-.1) (.6,.175) (.75,-.175) (.9,.1) (1.05,-.025) (1.2,0) (1.35,.0) };
                \draw [field, shift={(0,0)}, color=black, arrows = {-Stealth[inset=0, length=4pt, angle'=35]}] (0,0) -- (-0.1,0);
            \end{scope}
            \node at (1mm,6mm) {$\phv{f}^\swarrow$};
            \node at (-6mm,-1mm) {$\phv{f}^\nearrow_2$};
        \end{scope}
        \begin{scope}[shift={(-53mm,-5mm)}]
            \node[anchor=west] at (0,0mm) {spherical};
            \node[anchor=west] at (0,-3.3mm) {power};
            \node[anchor=west] at (0,-6mm) {waves};
        \end{scope}
        \draw[label_line] plot [smooth, samples=10, tension=0.9] coordinates {(-38mm,-5mm) (-36mm,-3.5mm) (-33mm,0mm)};
        \node[anchor=west] at (-20mm,19.5mm) {circuit-theoretic power waves};
        \draw[fill=white,white] (-.35,1.55) circle (.8mm);
        \draw[label_line] plot [smooth, samples=10, tension=0.9] coordinates {(-.4,1.78) (-.35,1.5) (-.3,1)};
    \end{tikzpicture}
    \caption{We consider the interaction of a radiating structure with (i) the circuit-theoretic power waves at its $M$ ports and (ii) the spherical power waves sufficiently far away in the radiating structure's far-field region. The former are characterized by the phasor vectors~$\phv{a}$ and~$\phv{b}$; and the latter by the angular spectrum~$\phv{f}^\swarrow$ of the incoming converging wave and the angular spectrum~$\phv{f}_2^\nearrow$ of the component of the outgoing diverging wave that is directly induced by the current density~$\phv{J}_2$.}
    \label{fig:setup_power}
    }
\end{figure}
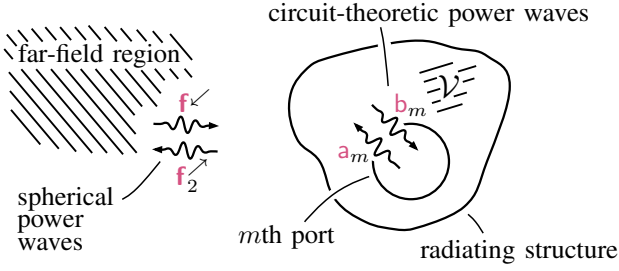

In addition to assuming finite size (cf.~\fref{asm:finite_size}), we restrict ourselves to (i) linear time-invariant (LTI) radiating structures (see~\fref{asm:linear_time_invariant}), (ii) with finitely many antenna ports (see~\fref{asm:finit_ports}), (iii) for which the induced current density~$\phv{J}_2$ is bounded by the maximal power that can interact with the radiating structure (see~\fref{asm:power_bound}). 
\begin{asm}[Linearity and Time-Invariance]\label{asm:linear_time_invariant}
    We assume that the radiating structure is an LTI system; that is, the operator mapping the externally impressed current density~$\phv{J}_1$ to the induced current density~$\phv{J}_2$ is linear and time-invariant.
\end{asm}
\begin{rem}
    It follows directly from~\fref{asm:linear_time_invariant} and from the linearity of Maxwell's equations that the operator mapping the externally impressed current density~$\phv{J}_1$ to the induced electromagnetic field~$(\phv{E}_2,\phv{H}_2)$ is also linear and time-invariant. 
\end{rem}

\begin{asm}[Finite Number of Ports]\label{asm:finit_ports}
    We assume that the interface between the radiating structure and the remainder of the device in which it is embedded---such as the RF front end---can be represented using a finite number of circuit-theoretic~ports. 
\end{asm}
\begin{asm}[Power-Bounded Current Response]\label{asm:power_bound}
    We assume the induced current density~$\phv{J}_2\in L^2({\mathcal{V}},\mathbb{C}^3)$ is upper-bounded~by 
    \begin{align}
        \|\phv{J}_2\|_{L^2(\mathcal{V},\mathbb{C}^3)}^2
        \leq
        C P_1,
    \end{align}
    where~\mbox{$C\in\mathbb{R}_{> 0}$} is a finite constant 
    and where~\mbox{$P_1\in\mathbb{R}_{\geq 0}$} is the maximal power that can interact with a given radiating structure for a given impressed current density~$\phv{J}_1\in L^2(\overline{\mathcal{V}},\mathbb{C}^3)$. 
\end{asm}
\begin{rem}
    If the ohmic resistance of all materials is lower-bounded by a positive ohmic resistance, then~\fref{asm:power_bound}~holds.
\end{rem}
As depicted in~\fref{fig:setup_power}, we consider a radiating structure's interactions (i)~through its \emph{ports} and (ii)~through its \emph{far-field region}. 
To characterize the former interaction, we rely on \emph{circuit-theoretic power waves} (see~\fref{defi:circuit_theoretic_power_waves}), and to characterize the latter interaction, we rely on \emph{spherical power waves} (see~\fref{defi:spherical_power_waves}). 
\begin{defi}[Circuit-Theoretic Power Waves]\label{defi:circuit_theoretic_power_waves}
    Given a system with $M\in\mathbb{Z}_{\geq 0}$ circuit-theoretic ports. 
    For $m\in [M]$, we define the \emph{circuit-theoretic power waves} traveling in and out of the system on the $m$th port as
    \begin{align}
        \label{eq:definition_power_waves}
        \phs{a}_m \triangleq \frac{1}{2\sqrt{R_0}}(\phs{v}_m+R_0 \phs{i}_m),
\,\,\,
        \phs{b}_m \triangleq \frac{1}{2\sqrt{R_0}}(\phs{v}_m-R_0\phs{i}_m). 
    \end{align}
    Here, $\phs{v}_m$ and $\phs{i}_m$ are the phasors of the voltage and current at port~$m$, respectively; and $R_0\in\mathbb{R}_{>0}$ is an arbitrary reference impedance (\SI{50}{\ohm} is a commonly-used choice).
\end{defi}
\begin{rem}\label{rem:charac_outgoing_port_waves}
    The interaction through the radiating structure's ports is fully characterized by \emph{circuit-theoretic power waves}. 
\end{rem}
\begin{defi}[Spherical Power Waves]\label{defi:spherical_power_waves}
    Given a system of finite size placed at the origin of the coordinate system. 
    Far away from the system, particularly as $r\rightarrow\infty$, the electric field relevant for the interaction with the system can be decomposed as follows\footnote{The normalization~$\sqrt{Z_\up{0}}$-factors ensure that~$\| \phv{f}^\swarrow \|_{L^2}^2$ and~$\| \phv{f}^\nearrow \|_{L^2}^2$ equal the total incoming and outgoing power, respectively.}~\cite[Eq.~5.12]{nieto_vesperinas_scattering_and_diffraction_in_physical_optics}:
    \begin{align}
        \label{eq:far_field_superposition}
        \!\lim_{r\rightarrow\infty}
        \frac{1}{\sqrt{Z_\up{0}}} r \phv{E}(r,\theta,\varphi)
        =
        \phv{f}^\swarrow(\theta,\varphi)e^{+jkr}
        +
        \phv{f}^\nearrow(\theta,\varphi)e^{-jkr}\!. 
    \end{align}
    Here~\mbox{$\phv{f}^\swarrow\!\!\in L^2(\Omega,\mathbb{C}^3)$} and~\mbox{$\phv{f}^\nearrow\!\!\in L^2(\Omega,\mathbb{C}^3)$} denote the angular spectra of the \emph{incoming, converging spherical power waves} and the \emph{outgoing, diverging spherical power waves}, respectively.
\end{defi}
\begin{rem}
    It follows from~\fref{asm:linear_time_invariant}, and from the linearity of Maxwell's equations, that one can further decompose the outgoing diverging spherical power waves as~\mbox{$\phv{f}^\nearrow=\phv{f}^\nearrow_1+\phv{f}^\nearrow_2$}, where~$\phv{f}^\nearrow_1,\phv{f}^\nearrow_2\in L^2(\Omega,\mathbb{C}^3)$ are the components directly induced by the current densities~$\phv{J}_1$ and~$\phv{J}_2$, respectively. 
\end{rem}
\begin{rem}\label{rem:charac_outgoing_far_field_waves}
    In the far-field region, the \emph{outgoing} EM waves that are induced in a radiating structure are fully characterized by the respective component of the angular spectrum of the outgoing divergent spherical power waves as~\cite[Sec.~4]{IEEE_standard_for_definitions_of_terms_for_antennas_2013}
    \begin{align}
    \label{eq:far_field_definition_approximation_out}
        \phv{E}_2(r,\theta,\varphi)
        &\approx
        \sqrt{Z_\up{0}}\phv{f}^\nearrow_2(\theta,\varphi)\frac{e^{-jkr}}{r}
        \\
        \phv{H}_2(r,\theta,\varphi)
        &\approx
        \frac{1}{\sqrt{Z_\up{0}}}
        \big(
        \hat{\vect{r}}\times
        \phv{f}^\nearrow_2(\theta,\varphi)
        \big)\frac{e^{-jkr}}{r}.
    \end{align}     
\end{rem}
\begin{rem}\label{rem:charac_incoming_far_field_waves}
    Sufficiently far away from the radiating structure, the \emph{incoming} EM waves that are sent toward a radiating structure are fully characterized by the angular spectrum of the incoming converging spherical power waves as~\cite[Eq.~5.12]{nieto_vesperinas_scattering_and_diffraction_in_physical_optics}
    \begin{align}
    \label{eq:far_field_definition_approximation_in}
        \phv{E}_1(r,\theta,\varphi)
        &\approx
        \sqrt{Z_\up{0}}\phv{f}^\swarrow(\theta,\varphi)\frac{e^{jkr}}{r}
        \\
        \phv{H}_1(r,\theta,\varphi)
        &\approx
        \frac{1}{\sqrt{Z_\up{0}}}
        \big(
        -\hat{\vect{r}}\times
        \phv{f}^\swarrow(\theta,\varphi)
        \big)\frac{e^{jkr}}{r}.
    \end{align}    
\end{rem}
From Remarks~\ref{rem:charac_outgoing_port_waves}-\ref{rem:charac_incoming_far_field_waves}, it follows that a radiating structure's port-and-far-field interaction\footnote{Specifically, we analyze the interaction of (i) EM waves that are incoming from sufficiently far away to be represented as convergent spherical waves and (ii) EM waves outgoing into the far-field region.} can be analyzed with the~\emph{effect} operator $\oper{T}:\mathbb{C}^M\times L^2(\Omega,\mathbb{C}^3)\rightarrow \mathbb{C}^M \times L^2(\Omega,\mathbb{C}^3)$ that we define as follows: 
\begin{align}
    \label{eq:defi_T}
    (\phv{b}, \phv{f}^\nearrow_2)
    \triangleq
    \oper{T} (\phv{a},\phv{f}^\swarrow).
\end{align}
In this work, we show that the effect operator $\oper{T}$ is of limited complexity, i.e., that it can be approximated arbitrarily well by a sequence of finite-rank operators (see~\fref{thm:finite_rank_approximation_far_field}). 
Furthermore, we provide an explicit sequence of finite-rank operators that converges to~$\oper{T}$ and an effective bandwidth beyond which the convergence is super-exponential (see~\fref{thm:vsh_approximation_far_field}).

\begin{thm}[Finite-Rank Representability of Far-Field Interactions]\label{thm:finite_rank_approximation_far_field}
    Given a radiating structure in free space for which Assumptions~\ref{asm:finite_size}-\ref{asm:power_bound} hold. 
    For any fixed analysis frequency~$f\in\mathbb{R}_{>0}$, let~$\oper{T}$ denote the effect operator introduced in~\fref{eq:defi_T}. 
    Then, the effect operator~$\oper{T}$ can be approximated arbitrarily well---in uniform operator topology---by finite-rank operators. 
\end{thm}
The proof of \fref{thm:finite_rank_approximation_far_field} is given in~\fref{sec:proof_thm_1}.

\begin{defi}[Scalar Spherical Harmonics]\label{defi:scalar_spherical_harmonics}
	For integers~\mbox{$\ell\geq 0$} and $|m|\leq\ell$, the (scalar) spherical harmonic of degree $\ell$ and order $m$ is the function $Y_\ell^m:\Omega\rightarrow\mathbb{C}$ defined by~\cite[Eq.~5.2(1)]{varshalovich_quantum_theory_of_angular_momentu}
	\begin{align}
		Y_\ell^m(\theta, \varphi)
		\triangleq
		\sqrt{\frac{2\ell+1}{4\pi}\frac{(\ell-m)!}{(\ell+m)!}}P_\ell^m(\cos(\theta))e^{jm\varphi},
	\end{align}
    where $P_\ell^m$ is the Legendre polynomial of degree~$\ell$ and order~$m$. 
\end{defi}
\begin{rem}
\label{rem:scalar_spherical_harmonics_orthonormal}
	The scalar spherical harmonics form a complete unitary basis of $L^2(\Omega, \mathbb{C})$~\cite[Sec.~2]{barrera_vector_spherical_harmonics_and_their_application_to_magnetostatics}. 
\end{rem}
\begin{defi}[Vector Spherical Harmonics (VSHs)]\label{defi:vector_spherical_harmonics}
    For integers~\mbox{$\ell\geq 0$} and $|m|\leq\ell$, the VSH of degree $\ell$ and order~$m$ are the functions with domain $\Omega$ and codomain $\mathbb{C}^3$ defined by~\cite[Sec.~3]{barrera_vector_spherical_harmonics_and_their_application_to_magnetostatics} 
	\begin{align}
		\vect{Y}_\ell^m(\uvect{r}) &\triangleq \uvect{r} Y_\ell^m(\uvect{r}), \\
		\vect{\Psi}_\ell^m(\uvect{r}) &\triangleq r\nabla Y_\ell^m(\uvect{r}),\label{eqn:curl_free_vsh} \\
		\vect{\Phi}_\ell^m(\uvect{r}) &\triangleq \vect{r} \times \nabla Y_\ell^m(\uvect{r}).
	\end{align}
\end{defi}
\begin{rem}
	\label{rem:vsh_orthogonal}
	The VSHs form a complete orthogonal basis of $L^2(\Omega, \mathbb{C}^3)$~\cite[Sec.~3]{barrera_vector_spherical_harmonics_and_their_application_to_magnetostatics}.
\end{rem}
\begin{defi}[VSH Projection Operator]\label{defi:approx_oper}
   For integers~\mbox{$L\geq 0$}, the VSH projection operator of degree~$L$ is the operator~\mbox{$\oper{P}_L:L^2(\Omega,\mathbb{C}^3)\rightarrow L^2(\Omega,\mathbb{C}^3)$} that maps vector fields to their orthogonal projection onto the subspace spanned by the VSHs of degree at most~$L$ 
   \begin{align}
        \mathcal{V}_L
        \triangleq 
        \spn
        \Big\{\vect{Y}_\ell^m, \vect{\Psi}_\ell^m, \vect{\Phi}_\ell^m\;\Big|\; 0 \leq |m| \leq \ell \leq L \Big\}.
    \end{align} 

\end{defi}
\begin{thm}[VSH Representability of Far-Field Interactions]\label{thm:vsh_approximation_far_field}
    Given a radiating structure in free space for which Assumptions~\ref{asm:finite_size}-\ref{asm:power_bound} hold. 
    For any fixed analysis frequency~$f\in\mathbb{R}_{>0}$ and any integer~\mbox{$L\geq 0$}, let~$\oper{T}$ denote the effect operator introduced in~\fref{eq:defi_T} and let~$\oper{T}_L$ be the composition\footnote{Specifically,~$\oper{T}_L$ is obtained by applying the VSH projection operator~$\oper{P}_L$ to the outgoing induced spherical power waves~$\phv{f}_2^\nearrow$, while leaving the outgoing circuit-theoretic power waves~$\phv{b}$ unchanged.} of the effect operator~$\oper{T}$ and the VSH projection operator~$\oper{P}_L$ of degree~$L$ (see~\fref{defi:approx_oper}).
    Furthermore, define the \emph{effective bandwidth} of this radiating structure as
    \begin{align}
        L_\up{B} 
        \triangleq
        \lceil ka \rceil,
    \end{align}
    where~$a$ denotes the smallest radius such that the radiating structure is completely contained in the ball~$\mathcal{B}_a$ defined in \fref{eq:ball}. 
    Then, 
    (i) the effect operator~$\oper{T}$ can be approximated arbitrarily well---in uniform operator topology---by the (finite-rank) operators~\mbox{$\{\oper{T}_L\}_{L\geq 0}$}, 
    and
    (ii) once~$L \geq L_\up{B}$, the approximation error $\|\oper{T} - \oper{T}_L\|_\up{op}$ exhibits super-exponential decay. 
    Specifically, there exists a function of the form~$\alpha e^{-\beta(L) L}$, where~$\alpha\in\mathbb{R}_{\geq0}$, and where~$\beta:\mathbb{Z}\rightarrow\mathbb{R}_{>0}$ is a monotonically increasing function for~$L\geq L_\up{B}$ satisfying~$\lim_{L\rightarrow\infty}
    \beta(L)= \infty$, so that 
    \begin{align}
        \label{eq:thm2}
        \| \oper{T} -\oper{T}_L \|_\up{op} \leq \alpha e^{-\beta(L) L}, \quad \forall L \geq L_\up{B}.
    \end{align}
\end{thm}
The proof of \fref{thm:vsh_approximation_far_field} is given in~\fref{sec:proof_thm_2}.

\begin{rem}
In~\fref{eq:K6}, we provide explicit expressions for both~$\alpha$ and~$\beta(L)$ such that~\fref{eq:thm2} is satisfied. 
\end{rem}


\section{Proof of Theorem 1}\label{sec:proof_thm_1}
\begin{proofw}
Before we begin with the actual proof, we emphasize that Assumptions~\ref{asm:finite_size}-\ref{asm:finit_ports} ensure that (i) the circuit-theoretic power waves at the ports are well defined and can be represented by the vectors~$\phv{a},\phv{b}\in\mathbb{C}^M$, where~$M$ is the number of ports, and (ii) that the spherical power waves around the radiating structure are likewise well defined and can be described by the angular spectrum functions~\mbox{$\phv{f}^\swarrow\!\!,\phv{f}^\nearrow_2\in L^2(\Omega,\mathbb{C}^3)$}. 
It follows directly from~\fref{asm:linear_time_invariant} and from the linearity of Maxwell's equations that the effect operator~$\oper{T}$ is linear. The quantities $\|\phv{a}\|_2^2$, $\|\phv{b}\|_2^2$, $\|\phv{f}^\swarrow\|^2_{L^2(\Omega,\mathbb{C}^3)}$, and~$\|\phv{f}^\nearrow_2\|^2_{L^2(\Omega,\mathbb{C}^3)}$ directly represent the power carried by their respective waves. Consequently, by energy conservation,~$\oper{T}$ is also bounded. 

The effect operator~$\oper{T}$ can be decomposed into two linear and bounded operators, \mbox{$\oper{T}^\phv{b}:\mathbb{C}^M\times L^2(\Omega,\mathbb{C}^3)\rightarrow \mathbb{C}^M$} and~\mbox{$\oper{T}^{\phv{f}_2^\nearrow}:\mathbb{C}^M\times L^2(\Omega,\mathbb{C}^3)\rightarrow L^2(\Omega,\mathbb{C}^3)$} as
\begin{align}
    (\phv{b},\phv{f}_2^\nearrow)
    =
    \oper{T} 
    \big(
        \phv{a}, \phv{f}^\swarrow
    \big) 
    =
    \Big(\oper{T}^\phv{b} 
    \big(
        \phv{a}, \phv{f}^\swarrow
    \big)
    ,
    \oper{T}^{\phv{f}_2^\nearrow}
    \big(
        \phv{a}, \phv{f}^\swarrow
    \big) 
    \Big).
\end{align}
The codomain of~\mbox{$\oper{T}^\phv{b}$} is finite-dimensional. Consequently,~$\oper{T}^\phv{b}$ is a finite-rank operator, and it is therefore sufficient to prove that~$\oper{T}^{\phv{f}_2^\nearrow}$ can be approximated by finite-rank operators. Since both the domain and codomain of~$\oper{T}^{\phv{f}_2^\nearrow}$ are Hilbert spaces, and all Hilbert spaces possess the \emph{approximation property}, it  suffices to show that~$\oper{T}^{\phv{f}_2^\nearrow}$ is a compact operator~\cite[Thm.~II-4.4]{conway_a_course_in_functional_analysis}.
The operator~$\oper{T}^{\phv{f}_2^\nearrow}$ can further be decomposed into the two linear operators~\mbox{$\oper{T}^{\phv{J}_2}_{(\phv{a},\phv{f}^\swarrow)}:\mathbb{C}^M\times L^2(\Omega,\mathbb{C}^3)\rightarrow L^2(\mathcal{V},\mathbb{C}^3)$} and~\mbox{$\oper{T}^{\phv{f}_2^\nearrow}_{\phv{J}_2}:L^2(\mathcal{V},\mathbb{C}^3)\rightarrow L^2(\Omega,\mathbb{C}^3)$} which are defined by
\begin{align}
    \label{eq:decomposed_operator}
    \phv{J}_2
    \triangleq
    \oper{T}^{\phv{J}_2}_{(\phv{a},\phv{f}^\swarrow)}
    \big(
        \phv{a}, \phv{f}^\swarrow
    \big),
    \quad
    \phv{f}_2^\nearrow
    \triangleq
    \oper{T}^{\phv{f}_2^\nearrow}_{\phv{J}_2}
    \phv{J}_2.
\end{align}
We will now show (i) that~$\oper{T}^{\phv{J}_2}_{(\phv{a},\phv{f}^\swarrow)}$ is bounded and (ii) that~$\oper{T}^{\phv{f}_2^\nearrow}_{\phv{J}_2}$ is compact. 
This, in turn, will imply that~$\oper{T}^{\phv{f}_2^\nearrow}$ is also compact; see~\cite[Prop.~VI-3.5]{conway_a_course_in_functional_analysis}.
The boundedness of~$\oper{T}^{\phv{J}_2}_{(\phv{a},\phv{f}^\swarrow)}$ follows from realizing that the sum~$\|\phv{a}\|_2^2+\|\phv{f}^\swarrow\|^2_{L^2(\Omega,\mathbb{C}^3)}$ is equal to the maximum available power~$P_1$ in~\fref{asm:power_bound}. 
Consequently, from~\fref{asm:power_bound} it follows that there exists a finite constant~$C\in\mathbb{R}$ so that 
\begin{align}
    \|\phv{J}_2\|_{L^2(\mathcal{V},\mathbb{C}^3)}
    \leq
    \sqrt{C P_1}
    =
    \sqrt{C}
    \sqrt{
    \|\phv{a}\|_2^2+\|\phv{f}^\swarrow\|^2_{L^2(\Omega,\mathbb{C}^3)}
    }.
\end{align}
It remains to be shown that~$\oper{T}^{\phv{f}_2^\nearrow}_{\phv{J}_2}$ is compact. 
For a given direction~$\hat{\vect{r}}$ the respective angular spectrum is given as 
\begin{align}
    \label{eq:A1}
    &\!\!\!\!\phv{f}^\nearrow_2(\hat{\vect{r}})
    =
    \frac{1}{\sqrt{Z_\up{0}}}
    \lim_{r\rightarrow\infty}
    r
    e^{j k r}
    \phv{E}_2(\hat{\vect{r}})
    \\
    \label{eq:A2}
    &=
    \frac{-1}{\sqrt{Z_\up{0}}}
    \lim_{r\rightarrow\infty}
    r
    e^{j k r}
    j\omega\mu_\up{0}\iiint_{\mathcal{V}}
    \mat{G}(\vect{r}-\vect{r}')
    \phv{J}_2(\vect{r}')
    \up{d}^3\vect{r}'
    \\
    \label{eq:A3}
    &=
    \frac{\omega \mu_0}{j\sqrt{Z_\up{0}}}
    \lim_{r\rightarrow\infty}
    r
    e^{j k r}
    \!\iiint_{\mathcal{V}}\!
        \frac{e^{-jkR}}{4\pi R}
        \left(
        \mat{I}-\hat{\vect{R}}\hat{\vect{R}}^\T
        \right)\!
    \phv{J}_2(\vect{r}')
    \up{d}^3\vect{r}'
    \\
    \label{eq:A4}
    &=
    \frac{\omega\mu_\up{0}}{j 4 \pi \sqrt{Z_\up{0}}}
    \left(
        \mat{I}_3-\hat{\vect{r}}\hat{\vect{r}}^\T
    \right)
    \lim_{r\rightarrow\infty}
    \iiint_{\mathcal{V}}
        e^{jk(r-R)}
    \phv{J}_2(\vect{r}')
    \up{d}^3\vect{r}'
    \\
    \label{eq:A5}
    &=
    \frac{\omega\mu_\up{0}}{j 4 \pi \sqrt{Z_\up{0}}}
    \left(
        \mat{I}_3-\hat{\vect{r}}\hat{\vect{r}}^\T
    \right)
    \iiint_{\mathcal{V}}
        e^{jk\hat{\vect{r}}^\T\vect{r}'}
    \phv{J}_2(\vect{r}')
    \up{d}^3\vect{r}',
\end{align}
where we use~$\vect{R}\triangleq\vect{r}-\vect{r}'$, $R\triangleq\|\vect{R}\|_2$, and~$\hat{\vect{R}}\triangleq\vect{R}/R$; 
and where~$\mat{G}:\mathbb{R}^3\rightarrow\mathbb{C}^{3\times3}$ is the dyadic Green's function. 
Here,~\fref{eq:A1} follows from~\fref{defi:spherical_power_waves};~\fref{eq:A2} and~\fref{eq:A3} follow from~\cite[Sec.~10.9]{paknys_applied_frequency_domain_electromagnetics};~\fref{eq:A4} follows from, as $r\to\infty$,~$\mat{I}_3-\hat{\vect{R}}\hat{\vect{R}}^\T$ converges to~$\mat{I}_3-\hat{\vect{r}}\hat{\vect{r}}^\T$, and $r/R$ converges to~$1$; and~\fref{eq:A5} follows from, as $r\to\infty$, $r - R $ converges to~$ \uvect{r}^\T\vect{r}'$.
From~\fref{eq:A5}, it follows that~$\oper{T}^{\phv{f}_2^\nearrow}_{\phv{J}_2}$ is an integral operator with a square-integrable kernel. 
Consequently, by~\cite[Prop.~II-4.7]{conway_a_course_in_functional_analysis}, the operator~$\oper{T}^{\phv{f}_2^\nearrow}_{\phv{J}_2}$ is compact.
\end{proofw}


\section{Proof of Theorem 2}\label{sec:proof_thm_2}
\begin{proofw}
We begin the proof by noting that property~(i) in \fref{thm:vsh_approximation_far_field} follows directly from property~(ii). It is therefore sufficient to prove property~(ii).
Furthermore, we ignore the circuit-theoretic power waves~$\phv{b}$ throughout the proof because $\oper{P}_L$ leaves the waves~$\phv{b}$ unchanged and, hence, they do not contribute to the approximation error~$\|\oper{T}-\oper{T}_L\|_\up{op}$. 
The proof is structured in the following three steps: 
In Step~I, we construct a sequence of finite-rank operators~$\big\{\oper{K}_L\big\}_{L\geq 0}$ and show that there exist a constant~$\alpha\in\mathbb{R}_{\geq0}$ and a function~\mbox{$\beta:\mathbb{Z}\rightarrow\mathbb{R}_{>0}$}, monotonically increasing for~$L\geq L_\up{B}$ and satisfying $\lim_{L\to\infty} \beta(L) = \infty$, such that
\begin{align}\label{eq:proof_step_1}
    \|\oper{T}-\oper{K}_L\|_\up{op}\leq\alpha e^{-\beta(L) L}.
\end{align}
In Step~II, we prove that for each~$L\geq 0$, the range of~$\oper{K}_L$ is contained in~$\mathcal{V}_L$. 
In Step~III, we invoke the orthogonality principle to show that, for all~$L\geq 0$, 
\begin{align}\label{eq:proof_step_3}
    \|\oper{T}-\oper{T}_L\|_\up{op}\leq \|\oper{T}-\oper{K}_L\|_\up{op}.
\end{align}
Combining~\fref{eq:proof_step_1} and~\fref{eq:proof_step_3} then yields that, for all~$L\geq L_\up{B}$,
\begin{align}
    \|\oper{T}-\oper{T}_L\|_\up{op}\leq\alpha e^{-\beta(L) L},
\end{align}
which proves~\fref{thm:vsh_approximation_far_field}. 

\subsection*{Step I: Finite-Rank Approximation:}
In analogy with~\fref{eq:decomposed_operator}, we decompose~$\oper{T}$ into the two linear operators \mbox{$\oper{T}^{\phv{J}_2}_{(\phv{a},\phv{f}^\swarrow)}:\mathbb{C}^M\times L^2(\Omega,\mathbb{C}^3)\rightarrow L^2(\mathcal{V},\mathbb{C}^3)$} and \mbox{$\oper{T}^{(\phv{b},\phv{f}_2^\nearrow)}_{\phv{J}_2}:L^2(\mathcal{V},\mathbb{C}^3)\rightarrow \mathbb{C}^M \times L^2(\Omega,\mathbb{C}^3)$}. 
For~$L\geq0$,
\begin{align}
    \label{eq:concat}
    \oper{K}_L
    &\triangleq
    \tilde{\oper{K}}_L
    \circ
    \oper{T}^{\phv{J}_2}_{(\phv{a},\phv{f}^\swarrow)},
\end{align}
    where we define~$\tilde{\oper{K}}_L:L^2(\mathcal{V},\mathbb{C}^3)\rightarrow \mathbb{C}^M \times L^2(\Omega,\mathbb{C}^3)$ as
\begin{align}
    \label{eq:F1}
    \left(\tilde{\oper{K}}_L
    \phv{J}_2\right)(\hat{\vect{r}})
    \triangleq \!
    \frac{\omega\mu_\up{0}}{j \sqrt{Z_\up{0}}} 
    \left(
        \mat{I}_3-\hat{\vect{r}}\hat{\vect{r}}^\T
    \right)
    \sum_{\ell=0}^{L-1} \sum_{m=-\ell}^\ell \! \vect{c}_{\ell, m} Y_\ell^m(\hat{\vect{r}})
\end{align} 
and where, for~$0\leq |m| \leq \ell$, we define the coefficient~\mbox{$\vect{c}_{\ell,m}\in\mathbb{C}^3$} as
\begin{align}
    \label{eq:defi_c}
    \vect{c}_{\ell,m}
    &\triangleq
    j^\ell \iiint_{\mathcal{V}}  \iota_\ell(kr') \overline{Y_\ell^m(\hat{\vect{r}}')} \phv{J}_2(\vect{r}') \up{d}^3\vect{r}'. 
\end{align} 
Here,~$Y_\ell^m$ is the scalar spherical harmonic of degree~$\ell$ and order~$m$ (see~\fref{defi:scalar_spherical_harmonics}) and~$\iota_\ell$ is the spherical Bessel functions of the first kind of degree~$\ell$.\footnote{To prevent confusion with the imaginary unit $j$, we deviate from the usual convention and use the letter $\iota$ to define the spherical Bessel functions.} 
The so-constructed~$\{\oper{K}_L\}_{L\geq 0}$ are (linear) finite-rank operators. 
By construction, it holds that 
\begin{align}
    \|\oper{T}-\oper{K}_L\|_\up{op}
    &=
    \left\|
    \left(
    \oper{T}_{\phv{J}_2}^{\phv{f}_2^\nearrow}-\tilde{\oper{K}}_L
    \right)
    \oper{T}_{(\phv{a},\phv{f}^\swarrow)}^{\phv{J}_2} 
    \right\|_\up{op}
    \\
    &\leq
    \left\|
    \oper{T}_{\phv{J}_2}^{\phv{f}_2^\nearrow}-\tilde{\oper{K}}_L 
    \right\|_\up{op}
    \left\|
    \oper{T}_{(\phv{a},\phv{f}^\swarrow)}^{\phv{J}_2} 
    \right\|_\up{op}, 
\end{align}
where $\left\|\oper{T}_{(\phv{a},\phv{f}^\swarrow)}^{\phv{J}_2}\right\|_\up{op}$ is finite because~$\oper{T}^{\phv{J}_2}_{(\phv{a},\phv{f}^\swarrow)}$ is bounded, as we showed in the proof of~\fref{thm:finite_rank_approximation_far_field}.
To bound $\left\|\oper{T}_{\phv{J}_2}^{\phv{f}_2^\nearrow}-\tilde{\oper{K}}_L \right\|_\up{op}$ from above, we now analyze $\oper{T}_{\phv{J}_2}^{\phv{f}_2^\nearrow}$.  
In the proof of~\fref{thm:finite_rank_approximation_far_field}, we derived that~$\oper{T}_{\phv{J}_2}^{\phv{f}_2^\nearrow}$ can be written as in~\fref{eq:A5}. 
We now apply the \emph{Jacobi–Anger expansion}~\cite[Eq.~2.46]{colton_cress_inverse_acoustic_and_electromagnetic_scattering_theory} to the kernel in~\eqref{eq:A5}:
\begin{align}
    \label{eq:C1}
	e^{jk\uvect{r}^\T\vect{r}'}
	=
	\sum_{\ell=0}^\infty j^\ell (2\ell+1) \iota_\ell(kr') P_\ell(\uvect{r}^\T\uvect{r}').
\end{align}
Here, $P_\ell$ is the Legendre polynomial of degree~$\ell$.  
After substituting this expression into~\fref{eq:A5}, we want to interchange the implicit limit with the integral. 
To justify this, we define the partial sums
\begin{align}
	S_L(\vect{r}')
	\triangleq
	\sum_{\ell=0}^L j^\ell (2\ell+1) \iota_\ell(kr') P_\ell(\uvect{r}^\T\uvect{r}').
\end{align}
Since the Jacobi-Anger expansion converges uniformly on compact subsets of~$\mathbb{R}^3$ (see~\cite[p.~37]{colton_cress_inverse_acoustic_and_electromagnetic_scattering_theory}), and each~$S_L$ is a finite sum of bounded functions, the sequence~$\{S_L\}_{L\geq0}$ is uniformly bounded on compact subsets of~$\mathbb{R}^3$ (see~\cite[Ex.~7.1]{rudin_principles_of_mathematical_analysis}).
Since $\mathcal{V}$ is bounded, it is contained in a compact subset of $\mathbb{R}^3$, and~$\{S_L\}_{L\geq0}$ is uniformly bounded on $\mathcal{V}$.
Therefore, for each~$L\geq 0$, each Cartesian basis vector $\vect{e}\in\{\vect{x}, \vect{y}, \vect{z}\}$, and each position~$\vect{r}'\in\mathcal{V}$, it holds that
\begin{align}
    \label{eq:B1}
	\Big|
    S_L(\vect{r}')[\phv{J}_2(\vect{r}')]_{\vect{e}} \Big|
	&\leq
	|S\sb{L}(\vect{r}')| \, \|\phv{J}_2(\vect{r}')\|_1\\
    \label{eq:B2}
	&\leq
	K \|\phv{J}_2(\vect{r}')\|_2,
\end{align}
for some finite $K\in\mathbb{R}$ that is independent of $L$. 
Here,~\fref{eq:B2} follows from the uniform boundedness of $\{S_L\}_{L\geq0}$ and the equivalence of norms on $\mathbb{C}^3$.
From~\fref{eq:B2} and~$\phv{J}_2 \in L^2(\mathcal{V}, \mathbb{C}^3)$ it follows that we can apply the Lebesgue dominated convergence theorem (see~\cite[Thm.~1.3.3]{evans_gariepy_measure_theory_and_fine_properties_of_functions}).
Consequently, after substituting~\fref{eq:C1} into~\fref{eq:A5}, we can interchange the implicit limit and the integral to arrive at 
\begin{align}
    &\!\!\!\iiint_{\mathcal{V}}
        e^{jk\hat{\vect{r}}^\T\vect{r}'}
    \phv{J}_2(\vect{r}')
    \up{d}^3\vect{r}'
    \nonumber
    \\
    \label{eq:D1}
    &=
    \sum_{\ell=0}^\infty
    \iiint_{\mathcal{V}}
    j^\ell (2\ell+1) \iota_\ell(kr') P_\ell(\uvect{r}^\T\uvect{r}')
    \phv{J}_2(\vect{r}')
    \up{d}^3\vect{r}'
    \\
    \label{eq:D2}
    &=
    4 \pi
    \sum_{\ell=0}^\infty \sum_{m=-\ell}^\ell j^\ell Y_\ell^m(\hat{\vect{r}}) \! \iiint_{\mathcal{V}} \! \iota_\ell(kr') \overline{Y_\ell^m(\hat{\vect{r}}')}\phv{J}_2(\vect{r}') \up{d}^3\vect{r}',
\end{align}
where~\fref{eq:D2} follows from the spherical harmonic addition theorem~\cite[Thm.~2.9]{colton_cress_inverse_acoustic_and_electromagnetic_scattering_theory}. 
Substituting~\eqref{eq:D2} into~\eqref{eq:A5} shows that~$\oper{T}_{\phv{J}_2}^{\phv{f}_2^\nearrow}$ may be written as 
\begin{align}
    \label{eq:F2}
    &\!\!\!\phv{f}^\nearrow_2(\hat{\vect{r}})
    =
    \frac{\omega\mu_\up{0}}{j\sqrt{Z_\up{0}}}
    \left(
        \mat{I}_3-\hat{\vect{r}}\hat{\vect{r}}^\T
    \right)
    \sum_{\ell=0}^\infty \sum_{m=-\ell}^\ell 
    \vect{c}_{\ell, m} Y_\ell^m(\hat{\vect{r}}).
\end{align}
We can now write the approximation error as\footnote{In~\fref{eq:E1} to \fref{eq:E3}, we omit an explicit specification of the corresponding Bochner spaces~\big($L^2(\mathcal{V},\mathbb{C}^3)$ and~$L^2(\Omega,\mathbb{C}^3)$\big) to keep notation simple.} 
\begin{align}
    &\!\!\!\left\|\oper{T}_{\phv{J}_2}^{\phv{f}_2^\nearrow}-\tilde{\oper{K}}_L \right\|_\up{op}
    \nonumber
    \\
    \label{eq:E1}
    &=
    \sup_{\|\phv{J}_2\|_{L^2}=1}
    \left\|
    \frac{\omega\mu_\up{0}}{j\sqrt{Z_\up{0}}}
    \left(
        \mat{I}_3-\hat{\vect{r}}\hat{\vect{r}}^\T
    \right)
    \sum_{\ell=L}^\infty \sum_{m=-\ell}^\ell 
    \vect{c}_{\ell, m} Y_\ell^m(\hat{\vect{r}})
    \right\|_{L^2}
    \\
    \label{eq:E2}
    &\leq
    \sup_{\|\phv{J}_2\|_{L^2}=1}
    \frac{\omega\mu_\up{0}}{\sqrt{Z_\up{0}}}
    \left\|
    \sum_{\ell=L}^\infty \sum_{m=-\ell}^\ell 
    \vect{c}_{\ell, m} Y_\ell^m(\hat{\vect{r}})
    \right\|_{L^2}
    \\
    \label{eq:E3}
    &=
    \sup_{\|\phv{J}_2\|_{L^2}=1}
    \frac{\omega\mu_\up{0}}{\sqrt{Z_\up{0}}}
    \sqrt{
    \sum_{\ell=L}^\infty \sum_{m=-\ell}^\ell 
    \left\|
    \vect{c}_{\ell, m}
    \right\|_{2}^2
    },
\end{align}
where~\fref{eq:E1} follows from~\fref{eq:F1} and~\fref{eq:F2}; 
\fref{eq:E2} follows because~$\mat{I}_3-\hat{\vect{r}}\hat{\vect{r}}^\T$ is an orthogonal projection; 
and~\fref{eq:E3} follows from~\fref{rem:scalar_spherical_harmonics_orthonormal} and Parseval's theorem. 

The coefficients in~\fref{eq:E3} can be bounded as follows\footnote{For points $(r,\theta,\varphi)$ that are not in the volume~$\mathcal{V}$, we define~$\phv{J}_2(\vect{r})=\vect{0}$.}:
\begin{align}
    \label{eq:G1}
    &\!\!\!\left\|
        \vect{c}_{\ell, m}
    \right\|_{2}^2
    =\left\|
        \iiint_{\mathcal{V}}  \iota_\ell(kr') \overline{Y_\ell^m(\hat{\vect{r}}')} \phv{J}_2(\vect{r}') \up{d}^3\vect{r}'
    \right\|_{2}^2
    \\
    \label{eq:G2}
    &=\left\|
        \int_0^a \iota_\ell(kr') \oiint_\Omega \overline{Y_\ell^m(\hat{\vect{r}}')} \phv{J}_2(\vect{r}') \up{d}^2\uvect{r}' r'^2 \up{d}r'
    \right\|_{2}^2
    \\
    \label{eq:G3}
    &\leq
    \int_0^a  
    \big(
    \iota_\ell(kr')
    \big)^2
    r'^2
    \up{d}r'
    \int_0^a
    \left\|
    \oiint_\Omega
    \overline{Y_\ell^m(\hat{\vect{r}}')} \phv{J}_2(\vect{r}')
    \up{d}^2\hat{\vect{r}}'
    \right\|_2^2
    r'^2
    \up{d}r'.
\end{align}
Here,~\fref{eq:G1} follows directly from~\fref{eq:defi_c}. 
In~\fref{eq:G2}, we use Fubini's theorem, which is applicable because (i)~$\phv{J}_2\in L^2(\mathcal{V},\mathbb{C}^3)$ and~$\mathcal{V}$ is bounded, implying that each component of $\phv{J}_2$ also belongs to~$L^1(\mathcal{V},\mathbb{C})$, and because (ii) both~$\iota_\ell$ and~$Y_\ell^m$ are bounded on the relevant bounded domains. 
Lastly, in~\fref{eq:G3}, we apply the Cauchy–Schwarz inequality. 
We now substitute~\fref{eq:G3} into the double sum in~\fref{eq:E3} to obtain 
\begin{align}
    &\!\!\!\sum_{\ell=L}^\infty \sum_{m=-\ell}^\ell 
    \left\|
    \vect{c}_{\ell, m}
    \right\|_{2}^2
    \leq 
    \sup_{\ell\geq L} 
    \int_0^a  
    \big(
    \iota_\ell(kr')
    \big)^2
    r'^2
    \up{d}r'
    \nonumber
    \\
    \label{eq:H1}
    &\times
    \sum_{\ell=L}^\infty \sum_{m=-\ell}^\ell 
    \int_0^a
    \left\|
    \oiint_\Omega
    \overline{Y_\ell^m(\hat{\vect{r}}')} \phv{J}_2(\vect{r}')
    \up{d}^2\hat{\vect{r}}'
    \right\|_2^2
    r'^2
    \up{d}r'. 
\end{align}
We now analyze the double-sum term in~\fref{eq:H1}, which can be bounded by 
\begin{align}
    &\!\!\!\sum_{\ell=L}^\infty \sum_{m=-\ell}^\ell 
    \int_0^a
    \left\|
    \oiint_\Omega
    \overline{Y_\ell^m(\hat{\vect{r}}')} \phv{J}_2(\vect{r}')
    \up{d}^2\hat{\vect{r}}'
    \right\|_2^2
    r'^2
    \up{d}r'
    \nonumber
    \\
    \label{eq:I1}
    &\leq
    \int_0^a
    \sum_{\ell=0}^\infty \sum_{m=-\ell}^\ell 
    \left\|
    \oiint_\Omega
    \overline{Y_\ell^m(\hat{\vect{r}}')} \phv{J}_2(\vect{r}')
    \up{d}^2\hat{\vect{r}}'
    \right\|_2^2
    r'^2
    \up{d}r'
    \\
    \label{eq:I2}
    &=
    \int_0^a
    \oiint_\Omega
    \left\|
    \phv{J}_2(\vect{r}')
    \right\|_2^2
    \up{d}^2\hat{\vect{r}}'
    r'^2
    \up{d}r'
    \\
    \label{eq:I3}
    &=
    \left\|
    \phv{J}_2
    \right\|_{L^2(\mathcal{V},\mathbb{C}^3)}^2,
\end{align}
where~\fref{eq:I1} follows from the monotone convergence theorem and~\fref{eq:I2} follows from~$\phv{J}_2\in L^2(\mathcal{V},\mathbb{C}^3)$, Fubini's theorem, and Parseval's theorem. 
We now insert~\fref{eq:I3} and~\fref{eq:H1} into~\fref{eq:E3} to obtain 
\begin{align}
    &\!\!\!\left\|\oper{T}_{\phv{J}_2}^{\phv{f}_2^\nearrow}-\tilde{\oper{K}}_L \right\|_\up{op}
    \nonumber
    \\
    \label{eq:J1}
    &\leq
    \sup_{\|\phv{J}_2\|_{L^2}=1}
    \frac{\omega\mu_\up{0}}{\sqrt{Z_\up{0}}}
    \sqrt{
        \sup_{\ell\geq L} 
        \int_0^a  
        \big(
        \iota_\ell(kr')
        \big)^2
        r'^2
        \up{d}r'
    \left\|
    \phv{J}_2
    \right\|_{L^2(\mathcal{V},\mathbb{C}^3)}^2
    }
    \\
    \label{eq:J2}
    &=
    \frac{\omega\mu_\up{0}}{\sqrt{Z_\up{0}}}
    \sqrt{
        \sup_{\ell\geq L} 
        \int_0^a  
        \big(
        \iota_\ell(kr')
        \big)^2
        r'^2
        \up{d}r'
    }
    \\
    \label{eq:J3}
    &=
    \frac{\omega\mu_\up{0}}{\sqrt{Z_\up{0}}}
    \sqrt{
        \sup_{\ell\geq L} 
        \frac{1}{k^3} \int_0^{ka} \left(\iota_\ell(u)\right)^2 u^2 \up{d}u
    }
    \\
    \label{eq:J4}
    &=
    \frac{\omega\mu_\up{0}}{\sqrt{Z_\up{0}}}
    \sqrt{
        \sup_{\ell\geq L} 
        \frac{\pi}{2k^3} \int_0^{ka} \left(I_{\ell+\frac{1}{2}}(u)\right)^2 u \,\up{d}u
    },
\end{align}
where~$I_\nu$ denotes the ordinary Bessel function of degree~\mbox{$\nu\in\mathbb{R}_{\geq 0}$}.\footnote{For consistent notation, we use the letter~$I$ to denote the ordinary Bessel functions.} 
Here,~\fref{eq:J3} follows from substituting~\mbox{$u\triangleq k r'$}; 
and~\fref{eq:J4}~follows from~\cite[Eq.~10.1.1]{abramowitz_stegun_handbook_of_mathematical_functions_with_formulas_graphs_and_mathematical_tables}. 
Next, we use the fact that for~$0\leq x\leq 1$ and~$\nu>0$, it holds that~\cite[Eq.~8]{siegel_an_inequality_involving_bessel_functions_of_argument_nearly_equal_to_their_order}
\begin{align}
    \label{eq:bound}
    (I_\nu(\nu x))^\frac{1}{\nu}\leq f(x)\triangleq\frac{x e^{\sqrt{1-x^2}}}{1+\sqrt{1-x^2}}.
\end{align} 
In the following, we choose~$x=\frac{u}{\ell+\frac{1}{2}}$ and~$\nu= \ell+\frac{1}{2}$. 
Consequently, for~\mbox{$L\geq L_\up{B}\triangleq\lceil k a\rceil$}, we can bound the approximation error as 
\begin{align}
    \label{eq:K1}
    &\!\!\!\left\|\oper{T}_{\phv{J}_2}^{\phv{f}_2^\nearrow}-\tilde{\oper{K}}_L \right\|_\up{op}
    \leq
    \frac{\omega\mu_\up{0}}{\sqrt{Z_\up{0}}}
    \sqrt{
        \sup_{\ell\geq L} 
        \frac{\pi}{2k^3} \int_0^{ka} 
            f(x)^{2\nu} 
        u\,\up{d}u
    }
    \\
    \label{eq:K2}
    &\leq
    \frac{\omega\mu_\up{0}}{\sqrt{Z_\up{0}}}
    \sqrt{
        \frac{\pi}{2k^3} \int_0^{ka} 
        f\!\left(\frac{u}{L+\frac{1}{2}}\right)^{2\left( L+\frac{1}{2} \right)} 
        u\,\up{d}u
    }
    \\
    \label{eq:K3}
    &\leq
    \frac{\omega\mu_\up{0}}{\sqrt{Z_\up{0}}}
    \sqrt{
        \frac{\pi}{2k^3} 
        \int_0^{ka} 
        u\,\up{d}u
        f\!\left(\frac{ka}{L+\frac{1}{2}}\right)^{2\left( L+\frac{1}{2} \right)} 
    }
    \\
    \label{eq:K4}
    &\leq
    \frac{\omega\mu_\up{0}\sqrt{\pi}a}{2\sqrt{k Z_\up{0}}}
    f \!
    \left(
        \frac{ka}{L+\frac{1}{2}}
    \right)^{ L+\frac{1}{2}}
    \\
    \label{eq:K5}
    &\leq
    \frac{\omega\mu_\up{0}\sqrt{\pi} a}{2\sqrt{k Z_\up{0}}}
    f \!
    \left(
        \frac{ka}{L+\frac{1}{2}}
    \right)^{L}
    \\
    \label{eq:K6}
    &=
    \underbrace{
    \frac{\omega\mu_\up{0}\sqrt{\pi}a}{2\sqrt{k Z_\up{0}}}
    }_{\alpha}
    \exp \!
    \left(
        \underbrace{
        \ln \!
        \left(
            f
            \left(
                \frac{ka}{L+\frac{1}{2}}
            \right)
        \right)    
        }_{=-\beta(L)}
        L
    \right)\!,
\end{align}
where~\fref{eq:K1} follows from~\fref{eq:bound}, and~\fref{eq:K2}-\fref{eq:K6} follow from the fact that on the interval~$0\leq x <1$,~$f(x)$ is monotonically increasing and~$0 \leq f(x) < 1$. 
Finally, we have~$\alpha \ge 0$, and since~$L \geq L_\up{B}$, it follows that~$\beta(L) > 0$ is a monotonically increasing function satisfying~$\lim_{L\rightarrow\infty} \beta(L)= \infty$, which concludes Step~I. 

\subsection*{Step II: VSH Subspace Containment:}
Because the VSHs form a complete basis of~$L^2(\Omega,\mathbb{C}^3)$ (see~\fref{rem:vsh_orthogonal}), it is sufficient to show that,  for $L' > L$ and $|{m'}|\leq L'$, the VSHs~$\vect{Y}_{L'}^{m'}$,~$\vect{\Psi}_{L'}^{m'}$, and~$\vect{\Phi}_{L'}^{m'}$ are orthogonal to any function that can be expressed in the form given on the right-hand side of~\fref{eq:F1}.
To this end, we note that $\vect{Y}_{L'}^{m'}$ is purely radial, whereas the right-hand side of~\fref{eq:F1} is purely transversal; consequently, they are orthogonal.
Next, we fix arbitrary coefficients~$\{\vect{c}_{\ell,m}\}_{\ell,m}$. 
We use~$\vect{X}_{L'}^{m'}$ as a placeholder for~$\vect{\Psi}_{L'}^{m'}$ or~$\vect{\Phi}_{L'}^{m'}$ and analyze the inner product 
\begin{align}
    \label{eq:Y1}
    &\!\!\!
    \left\langle
    \left(
        \mat{I}_3-\hat{\vect{r}}\hat{\vect{r}}^\T
    \right)
    \sum_{\ell=0}^{L-1} \sum_{m=-\ell}^\ell \! \vect{c}_{\ell, m} Y_\ell^m(\hat{\vect{r}})
    ,
    \vect{X}_{L'}^{m'}
    \right\rangle_{L^2(\Omega,\mathbb{C}^3)}
    \nonumber\\
    &=
    \oiint_\Omega \left( \left(
        \mat{I}_3-\hat{\vect{r}}\hat{\vect{r}}^\T
    \right)
    \sum_{\ell=0}^{L-1} \sum_{m=-\ell}^\ell \! \vect{c}_{\ell, m} Y_\ell^m(\hat{\vect{r}})
    \right)^\He
    \vect{X}_{L'}^{m'} \up{d}^2\uvect{r}
    \\
    \label{eq:Y2}
    &=
    \oiint_\Omega \left(
    \sum_{\ell=0}^{L-1} \sum_{m=-\ell}^\ell \! \vect{c}_{\ell, m} Y_\ell^m(\hat{\vect{r}})
    \right)^\He
    \left(
        \mat{I}_3-\hat{\vect{r}}\hat{\vect{r}}^\T
    \right)
    \vect{X}_{L'}^{m'} \up{d}^2\uvect{r}
    \\
    &=
    \left\langle
    \sum_{\ell=0}^{L-1} \sum_{m=-\ell}^\ell \! \vect{c}_{\ell, m} Y_\ell^m(\hat{\vect{r}})
    ,
    \vect{X}_{L'}^{m'}
    \right\rangle_{L^2(\Omega,\mathbb{C}^3)}\hspace{-10mm},
    \label{eq:Y3}
\end{align}
where in~\fref{eq:Y3} we used the transversality of~$\vect{\Psi}_{L'}^{m'}$ and~$\vect{\Phi}_{L'}^{m'}$. 
From~\cite[Eq.~7.3(3) and Eq.~7.3(9)]{varshalovich_quantum_theory_of_angular_momentu} we conclude that~$\vect{X}_{L'}^{m'}$ can be written as
\begin{align}
    \vect{X}_{L'}^{m'}
    =
    \sum_{\ell={L'}-1}^{{L'}+1}\sum_{m=-\ell}^\ell {\vect{c}'}_{\ell,m} Y_\ell^m(\hat{\vect{r}}),
\end{align}
where ${\vect{c}'}_{\ell,m}\in\mathbb{C}^3$.
From this representation, it follows that $\vect{X}_{L'}^{m'}$ involves scalar spherical harmonics of minimum degree ${L'}-1\geq L$. 
However, the sum in~\fref{eq:Y3} contains only scalar spherical harmonics of degrees at most $L-1$, so there is no overlap in degree.
By the orthogonality of scalar spherical harmonics (see~\fref{rem:scalar_spherical_harmonics_orthonormal}), it follows that $\vect{X}_{L'}^{m'}$ is orthogonal to any function of the form given on the right-hand side of~\fref{eq:F1}.
We conclude that for~$L\geq0$, it holds that
\begin{align}
    \label{eq:range_in_subset}
    \range\{\oper{K}_L\}\subset \mathcal{V}_L.
\end{align}

\subsection*{Step III: Orthogonality Principle:}
For any~$L\geq 0$ it holds that
\begin{align}
    \label{eq:N1}
    &\!\!\!\|\oper{T}-\oper{T}_L\|_\up{op}
    =
    \sup_{\|\phv{J}_2\|_{L^2}=1}
    \left\|
    \left(
    \oper{T}-\oper{T}_L
    \right)
    \phv{J}_2
    \right\|_{L^2}
    \\
    \label{eq:N2}
    &=
    \sup_{\|\phv{J}_2\|_{L^2}=1}
    \left\|
    \left(
    \oper{I}-\oper{P}_L
    \right)
    \oper{T}
    \phv{J}_2
    \right\|_{L^2}
    \\
    \label{eq:N3}
    &=
    \sup_{\|\phv{J}_2\|_{L^2}=1}
    \left\|
    \left(
    \oper{I}-\oper{P}_L
    \right)
    \left(
        \oper{T}
        -
        \oper{K}_L
        +
        \oper{K}_L
    \right)
    \phv{J}_2
    \right\|_{L^2}
    \\
    \label{eq:N4}
    &=
    \sup_{\|\phv{J}_2\|_{L^2}=1}
    \left\|
    \left(
    \oper{I}-\oper{P}_L
    \right)
    \left(
        \oper{T}
        -
        \oper{K}_L
    \right)
    \phv{J}_2
    \right\|_{L^2}
    \\
    \label{eq:N5}
    &\leq
    \sup_{\|\phv{J}_2\|_{L^2}=1}
    \left\|
    \left(
        \oper{T}
        -
        \oper{K}_L
    \right)
    \phv{J}_2
    \right\|_{L^2}
    \\
    \label{eq:N6}
    &= \|\oper{T}-\oper{K}_L\|_\up{op},
\end{align}
where~$\oper{I}$ is the identity operator. 
Here,~\fref{eq:N4} follows directly from~\fref{eq:range_in_subset}
and~\fref{eq:N5} follows from the fact that~$\oper{P}_L$ is an orthogonal projection. 
Finally, as discussed at the beginning of \fref{sec:proof_thm_1}, the result in~\fref{eq:N6} concludes the proof. 
\end{proofw}

\section{Conclusions}\label{sec:conclusions}
We have shown that the electromagnetic far-field interaction of a wide range of radiating structures can be approximated arbitrarily well by a sequence of finite-rank operators (see~\fref{thm:finite_rank_approximation_far_field}). 
Furthermore, we have introduced a vector-spherical-harmonics-based method to construct such a sequence of finite-rank operators. %
    In particular, we have demonstrated that the approximation error exhibits super-exponential decay once the rank exceeds an effective bandwidth (see~\fref{thm:vsh_approximation_far_field}). 
Our results imply that the far-field interaction of a wide range of wireless systems can be modeled with limited complexity---using finitely many parameters---up to arbitrary accuracy.
This result further strengthens our justification in~\cite[Rem.~9]{stutz_schwan_studer_efficient_and_physically_consistent_modeling_of_reconfigurable_electromagnetic_structures} that the scattering operator defined in~\cite[Eq.~21]{stutz_schwan_studer_efficient_and_physically_consistent_modeling_of_reconfigurable_electromagnetic_structures} can be represented accurately (up to arbitrary precision) using a finite number of model~parameters. 

\balance
\bstctlcite{IEEEexample:BSTcontrol} 
\bibliographystyle{IEEEtran}
\bibliography{bib/publishers,bib/journals_proceedings_ect,bib/library}
\balance

\end{document}